\documentclass[12pt]{article}

\usepackage{axodraw}
\usepackage{epsf}
\usepackage{rotating}
\usepackage{epsfig}
\usepackage{slashbox}
\usepackage[dvips]{color}

\newcommand{\beq}{\begin{equation}}
\newcommand{\eeq}{\end{equation}}
\newcommand{\bea}{\begin{eqnarray}}
\newcommand{\eea}{\end{eqnarray}}
\newcommand{\bed}{\begin{displaymath}}
\newcommand{\eed}{\end{displaymath}}

\newcommand{\tgb}{{\rm tg}\beta}

\catcode`\@=11

\def\@citex[#1]#2{\if@filesw\immediate\write\@auxout{\string\citation{#2}}\fi
  \def\@citea{}\@cite{\@for\@citeb:=#2\do
    {\@citea\def\@citea{,\penalty\@m}\@ifundefined
       {b@\@citeb}{{\bf ?}\@warning
       {Citation `\@citeb' on page \thepage \space undefined}}%
\hbox{\csname b@\@citeb\endcsname}}}{#1}}

\def\citer{\@ifnextchar [{\@tempswatrue\@citexr}{\@tempswafalse\@citexr[]}}
\def\@citexr[#1]#2{\if@filesw\immediate\write\@auxout{\string\citation{#2}}\fi
  \def\@citea{}\@cite{\@for\@citeb:=#2\do
    {\@citea\def\@citea{--\penalty\@m}\@ifundefined
       {b@\@citeb}{{\bf ?}\@warning
       {Citation `\@citeb' on page \thepage \space undefined}}%
\hbox{\csname b@\@citeb\endcsname}}}{#1}}
\catcode`\@=12

\begin{document}

\renewcommand{\thefootnote}{\fnsymbol{footnote}}
\setcounter{page}{0}

\begin{titlepage}

\vskip-1.0cm

\begin{flushright}
PSI--PR--15--08
\end{flushright}

\begin{center}
{\large \sc Higgs Boson Production via Gluon Fusion:} \\[0.3cm]
{\large \sc Soft-Gluon Resummation including Mass Effects}\\
\end{center}

\vskip 1.cm
\begin{center}
{\sc Timo Schmidt$^{1,2}$\footnote{present address:
Albert-Ludwigs-Universit\"at Freiburg, Physikalisches Institut, D--79104
Freiburg, Germany} and Michael Spira$^1$}

\vskip 0.8cm

\begin{small} 
{\it \small
$^1$ Paul Scherrer Institut, CH--5232 Villigen PSI, Switzerland \\
$^2$ Physik Institut, Z\"urich University, CH--8057
Z\"urich, Switzerland}
\end{small}
\end{center}

\vskip 2cm

\begin{abstract}
\noindent
We analyze soft and collinear gluon resummation effects at the N$^3$LL
level for Standard Model Higgs boson production via gluon fusion $gg\to
H$ and the neutral scalar and pseudoscalar Higgs bosons of the minimal
supersymmetric extension at the N$^3$LL and NNLL level, respectively. We
introduce refinements in the treatment of quark mass effects and
subleading collinear gluon effects within the resummation. Soft and
collinear gluon resummation effects amount to up to about 5\% beyond the
fixed-order results for scalar and pseudoscalar Higgs boson production.
\end{abstract}

\end{titlepage}

\renewcommand{\thefootnote}{\arabic{footnote}}

\setcounter{footnote}{0}

\section{Introduction}
The Standard Model (SM) of elementary particle physics has been
established as a very successful theory that describes the properties
and strong, weak and electromagnetic interactions of all known
elementary particles \cite{gsw}.  The recent discovery of the Higgs
boson with a mass of about 125 GeV \cite{discovery} completed the
particle content of the SM and established the Higgs mechanism for
electroweak symmetry breaking \cite{hi64}. Its couplings to SM gauge
bosons, i.e.~to $ZZ$, $W^+W^-$, and fermion pairs ($\tau$ leptons and
bottom quarks) as well as the loop-induced couplings to gluon and photon
pairs, have been measured with accuracies of 10--50\%. All measured
couplings are in agreement with the SM predictions within their
uncertainties \cite{couplings}. In addition there are very strong
indications that the newly discovered boson carries zero spin and
positive ${\cal CP}$-parity \cite{cpspin}. Its discovery is of vital
importance for the mathematical consistency of the SM and the success of
the predictions for the precision electroweak observables which are in
striking agreement with measurements at LEP and SLC \cite{elwpo}. Based
on the present situation it is highly relevant to test the properties of
the discovered particle in more detail. The measured inclusive
production and decay rates are in agreement with the theoretical
predictions within the corresponding uncertainties.

At hadron colliders as the LHC neutral Higgs bosons of the Standard
Model are copiously produced by the gluon fusion process $gg\to H$,
which is mediated by top and to a lesser extent bottom and charm quark
loops.  Due to the large top Yukawa coupling and the large gluon
luminosities gluon fusion comprises the dominant Higgs boson production
mechanism for the SM \cite{gghlo}.

In the past the NLO QCD corrections to the top and bottom/charm quark
loops have been calculated \cite{gghnlolim,gghnlo,gghnlo0,gghnlosq0}.
They increase the cross sections by up to 90\%. The full quark and Higgs
mass dependencies for the quark loops have been included
\cite{gghnlo,gghnlo0,gghnlosq0}. The scale dependence decreased from
${\cal O}(100\%)$ to $\sim 20\%$ at NLO. The NLO results indicated that
the limit of heavy top quarks provides a reliable approximation of the
full relative QCD corrections within $\sim 5\%$ for a Higgs mass of 125
GeV \cite{gghnlo,limit}. In this limit the cross section factorizes into
a part originating from an effective Lagrangian derived by integrating
out the top quark supplemented by gluonic and light quark corrections
within the effective low-energy theory \cite{gghnlolim,let}. Within this
approach the NNLO QCD corrections have been calculated leading to a
further increase of the cross section by 20--30\%
\cite{gghnnlo,gghnnlosv,gghannlo} and a further reduction of the scale
dependence to $\sim 10\%$. Mass effects beyond the heavy top
approximation have been studied at NNLO in a heavy top mass expansion
\cite{gghnnlomt}.  These mass effects amount to less than a per cent in
the relative QCD corrections below the $t\bar t$-threshold. The NNLO
results have been improved by a soft and collinear gluon resummation at
the NNLL level \cite{gghnnll,gghnnll0} that add another 5--10\% beyond
NNLO to the total Higgs production cross section. In addition the small
quark mass effects have been included in the soft-gluon-resummed result
at NLL \cite{gghnnllmt}.  Very recently the N$^3$LO corrections have
been computed in the limit of heavy top quarks first approximately
deriving the leading terms in a threshold expansion \cite{gghn3loapprox}
and then completely \cite{gghn3lo,gghn3lo2}. These increase the cross
section by only a few per cent and reduce the scale dependence to a
level of about 5\%.  They coincide with the soft+collinear gluon
approximation at N$^3$LO at the level of a few per cent
\cite{gghn3ll1,gghn3ll2,gghn3ll3}.  This signalizes that soft gluon
effects are less important at N$^3$LL than at NNLL level thus
underlining a proper perturbative reliability of the N$^3$LO and N$^3$LL
results. These results have been completed by the calculation of the
full electroweak corrections at NLO \cite{gghelw} and beyond
\cite{gghelw1} that provide a further increase of the cross section by
about 5\%.

One of the most attractive extensions of the SM is the minimal
supersymmetric SM (MSSM) that requires the introduction of two Higgs
doublets and thus predicts the existence of five elementary Higgs
bosons, two neutral CP-even $h,H$, one neutral CP-odd $A$ and two
charged ones $H^\pm$ \cite{twoiso}. A LO the Higgs sector is described
by two independent input parameters that in the case of a real MSSM are
usually chosen to be the pseudoscalar Higgs mass $M_A$ and the parameter
$\tgb$ defined as the ratio of the two CP-even vacuum expectation
values.  Higher-order corrections to the MSSM Higgs masses and couplings
turned out to be large with the dominant piece originating from
contributions induced by the large top Yukawa coupling \cite{mssmrad}.
Including all relevant corrections up to the three-loop level the mass
of the light scalar Higgs boson $h$ is bound to be smaller than $\sim
135$ GeV \cite{mh3l}. The light scalar Higgs boson becomes SM-Higgs like
close to its upper mass bound thus allowing for the possibility of the
Higgs boson found at the LHC to be identified with the light scalar $h$.
Global fits to the MSSM leave a small region within the MSSM parameter
space where the discovered Higgs particle could also be the heavy scalar
$H$ \cite{heinlein} but this region is disfavoured by LHC data
\cite{heavyhiggs}. The Higgs couplings to intermediate gauge bosons and
fermions are modified by additional factors shown in Table \ref{tb:hcoup}
that only depend on the mixing angles $\alpha$ and $\beta$ where $\alpha$
denotes the mixing angle between the neutral CP-even Higgs states.
\begin{table}[hbt]
\renewcommand{\arraystretch}{1.5}
\begin{center}
\begin{tabular}{|lc||ccc|} \hline
\multicolumn{2}{|c||}{$\phi$} & $g^\phi_u$ & $g^\phi_d$ &  $g^\phi_V$ \\
\hline \hline
SM~ & $H$ & 1 & 1 & 1 \\ \hline
MSSM~ & $h$ & $\cos\alpha/\sin\beta$ & $-\sin\alpha/\cos\beta$ &
$\sin(\beta-\alpha)$ \\ & $H$ & $\sin\alpha/\sin\beta$ &
$\cos\alpha/\cos\beta$ & $\cos(\beta-\alpha)$ \\
& $A$ & $ 1/\tgb$ & $\tgb$ & 0 \\ \hline
\end{tabular}
\renewcommand{\arraystretch}{1.2}
\caption[]{\label{tb:hcoup} \it Higgs couplings in the MSSM to fermions
and gauge bosons [$V=W,Z$] relative to the SM couplings.}
\end{center}
\end{table}
For large values of $\tgb$ the down-type Yukawa couplings are strongly
enhanced, while the up-type Yukawa couplings are suppressed, unless the
light (heavy) scalar Higgs mass ranges at its upper (lower) bound, where
the couplings become SM-sized.

The dominant Higgs boson production mechanisms for small and moderate
values of $\tgb$ are the gluon fusion processes $gg\to h,H,A$ that are
mediated by top and bottom and in addition stop and sbottom loops in the
case of the scalar Higgs bosons $h,H$. For large $\tgb$ the leading role
is taken over by Higgs radiation off bottom quarks due to the strongly
enhanced bottom Yukawa couplings \cite{habil}. In the past the NLO QCD
corrections for pseudoscalar Higgs boson production have been derived in
the limit of heavy top quarks \cite{gganlolim} and later including the
full Higgs and quark mass dependencies \cite{gghnlo,gganlo}. They are
large ranging up to about 100\%. In the limit of heavy top quarks the
calculation of the SM Higgs boson at NNLO QCD has been extended to the
pseudoscalar case, too \cite{gghannlo,ggannlo} inducing a moderate
increase of the total cross section by about 20--30\% as in the SM Higgs
case. However, the limit of heavy top quarks is only applicable for the
pure top loop contributions so that for bottom-loop dominance for large
values of $\tgb$ we are left with NLO accuracy. For the case of top-loop
dominance, i.e.~for small and moderate values of $\tgb$ within the MSSM,
the soft-gluon resummation has been extended to the NNLL level for the
pseudoscalar Higgs boson \cite{ggannll} resulting in corrections of
${\cal O}(10\%)$ beyond NNLO. For the light and heavy scalar MSSM Higgs
bosons the NLO QCD corrections to the squark loops have been calculated
in the approximate limit of heavy squarks \cite{gghnlosqlim} yielding
NLO corrections close to 100\% as for the quark loops. The
next-to-leading order computation for the squark loops including the
full mass dependencies has been performed later
\cite{gghnlosq0,gghnlosq} for the pure QCD corrections. The genuine
supersymmetric QCD corrections have been derived in
Ref.~\cite{gghnlosqcd} for large supersymmetric particle masses,
i.e.~the full Higgs mass dependence has not been taken into account. For
gluino masses much larger than the stop and top masses the results of
\cite{gghnlosqcd} develop a logarithmic singularity in the gluino mass
that seems to contradict the Appelquist--Carazzone decoupling theorem
\cite{appcar} at first sight.  This problem has been solved by a
detailed renormalization group analysis with the corresponding
decoupling techniques for heavy particles \cite{heavygluino}.
The full supersymmetric QCD corrections including all mass dependencies
has been completed in Refs.~\cite{gghnlofull}. The genuine SUSY--QCD
corrections at NLO are large, too, modifying the total gluon-fusion
cross sections by up to $\sim 100\%$ depending on the MSSM parameters of
the Higgs and stop/sbottom sectors.

In this work we will present the soft+virtual+collinear gluon resummed
results at N$^3$LL for scalar Higgs boson production via gluon fusion
including a consistent treatment of top and bottom mass effects with an
extension to the inclusion of additional subleading collinear gluon
effects. For the pseudoscalar Higgs boson we will present the analogous
results up to the NNLL level, since the fixed-order N$^3$LO result does
not exist for this case. The paper is organized as follows.  In Section
2 we will describe our method for the scalar and pseudoscalar Higgs
bosons of the SM and MSSM.  In Section 3 we will present results and
discuss the comparison with previous calculations. In Section 4 we will
finally present our conclusions.

\section{SM Higgs Boson Production}
Higgs boson production via gluon fusion is mediated by top and to a
lesser extent bottom and charm triangle loops at LO. Following the
notation of Ref. \cite{gghnnll} the production cross section of scalar
Higgs boson production via gluon fusion $gg\to H$ can be cast into the
form
\begin{eqnarray}
\sigma (s,M_H^2) & = & \sigma_0 \sum_{ij} \int_{\tau_H}^1 d\tau \frac{d{\cal
L}^{ij}}{d\tau}
\int_0^1 dz~z~G_{ij}\left( z;\alpha_s(\mu_R^2),\frac{M_H^2}{\mu_R^2};
\frac{M_H^2}{\mu_F^2} \right) \delta\left(z-\frac{\tau_H}{\tau}\right)
\nonumber \\
\sigma_0 & = & \frac{G_F}{288\sqrt{2}\pi} \left| \sum_Q
A_Q(\tau_Q)\right|^2 \nonumber \\
A_Q(\tau) & = & \frac{3}{2}\tau[1+(1-\tau) f(\tau)] \nonumber \\
f(\tau) & = & \left\{ \begin{array}{ll}
\displaystyle \arcsin^2 \frac{1}{\sqrt{\tau}} & \tau \ge 1 \\
\displaystyle - \frac{1}{4} \left[ \log \frac{1+\sqrt{1-\tau}}
{1-\sqrt{1-\tau}} - i\pi \right]^2 & \tau < 1
\end{array} \right.
\label{eq:ggh}
\end{eqnarray}
where the sum over $i,j$ runs over all contributing initial state,
i.e.~only gluons at LO. The term ${\cal L}^{ij}$ denotes the
corresponding parton-parton luminosity, $M_H$ the Higgs boson mass,
$\mu_{R/F}$ the renormalization/factorization scale, $\alpha_s$ the
strong coupling constant and $G_F$ the Fermi constant. The variable
$\tau_Q$ is defined as
\begin{equation}
\tau_Q = 4\frac{m_Q^2}{M_H^2}
\end{equation}
where $m_Q$ is the corresponding loop quark mass. The lower integration
bound is given by $\tau_H = M_H^2/s$. The integration kernel $G_{ij}$
can be computed as a perturbative expansion in the strong coupling
constant,
\begin{equation}
G_{ij} \left( z;\alpha_s(\mu_R^2),\frac{M_H^2}{\mu_R^2};
\frac{M_H^2}{\mu_F^2} \right) = \alpha_s^2(\mu_R^2) \sum_{n=0}^{\infty}
\left( \frac{\alpha_s(\mu_R^2)}{\pi} \right)^n G_{ij}^{(n)} \left(
z;\frac{M_H^2}{\mu_R^2}; \frac{M_H^2}{\mu_F^2} \right)
\end{equation}
where the LO and NLO expressions are given by
\begin{eqnarray}
G_{ij}^{(0)} (z) & = & \delta_{ig} \delta_{jg} (1-z) \nonumber \\
G_{gg}^{(1)} \left( z;\frac{M_H^2}{\mu_R^2}; \frac{M_H^2}{\mu_F^2}
\right) & = & \delta(1-z) \left\{ c_H(\tau_Q) + 6\zeta_2 +
\frac{33-2N_F}{6} \log \frac{\mu_R^2}{\mu_F^2} \right\} + 12 {\cal
D}_1(z) \nonumber \\
& + & 6 {\cal D}_0(z)\log\frac{M_H^2}{\mu_F^2} + P_{gg}^{reg} (z) \log
\frac{(1-z)^2 M_H^2}{z\mu_F^2} -6\frac{\log z}{1-z} + d^H_{gg}(z,\tau_Q)
\nonumber \\
G_{gq}^{(1)} \left( z;\frac{M_H^2}{\mu_R^2}; \frac{M_H^2}{\mu_F^2}
\right) & = & \frac{1}{2} P_{gq}(z) \log \frac{(1-z)^2 M_H^2}{z\mu_F^2}
+ d^H_{gq}(z,\tau_Q) \nonumber \\
G_{q\bar q}^{(1)} \left( z;\frac{M_H^2}{\mu_R^2}; \frac{M_H^2}{\mu_F^2}
\right) & = & d^H_{q\bar q}(z,\tau_Q)
\label{eq:gghnlo}
\end{eqnarray}
with the NLO quark mass dependence contained in the functions $c_H$ and
$d^H_{ij}~(ij=gg, gq, q\bar q)$. Explicit results for these functions
can be found in \cite{gghnlo}\footnote{Note that the results of
\cite{gghnlo} have to be divided by $z$ due to our different
normalization of the functions $d_{ij}^H(z,\tau_Q)$ $(ij = gg,gq,q\bar
q)$.}. In the limit of heavy top quarks they
approach the following simple expressions,
\begin{eqnarray}
c_H(\tau_Q) & \to & \frac{11}{2} \nonumber \\
d^H_{gg}(z,\tau_Q) & \to & -\frac{11}{2} \frac{(1-z)^3}{z} \nonumber \\
d^H_{gq}(z,\tau_Q) & \to & \frac{2}{3}z - \frac{(1-z)^2}{z} \nonumber \\
d^H_{q\bar q}(z,\tau_Q) & \to & \frac{32}{27} \frac{(1-z)^3}{z}
\label{eq:dijlim}
\end{eqnarray}
For the NLO results presented above we have used the notation
($i=0,1,\ldots$)
\begin{equation}
{\cal D}_i(z) = \left( \frac{\log^i(1-z)}{1-z} \right)_+
\end{equation}
for the plus distributions and $\zeta_2 = \pi^2/6$.  The
Altarelli--Parisi splitting kernels are given by \cite{altpar}
\begin{eqnarray}
P_{gg}(z) & = & 6 {\cal D}_0(z) + P_{gg}^{reg} (z) \nonumber \\
P_{gg}^{reg}(z) & = & 6 \left[ \frac{1}{z} - 2 + z(1-z) \right]
\nonumber \\
P_{gq}(z) & = & \frac{4}{3} \frac{1+(1-z)^2}{z}
\end{eqnarray}
The results of the NNLO pieces $G_{ij}^{(2)}$ of the coefficient
function can be extracted from Refs.~\cite{gghnnlo,gghannlo} and the
Mellin transforms of their leading soft+virtual contributions are given
explicitly in Ref.~\cite{gghnnll}.

\subsection{Soft and collinear Gluon Resummation}
In this work we will resum soft and collinear gluon effects up to all
orders in the perturbative expansion. This will be performed
systematically in Mellin space. The Mellin moment of the gluon-fusion
cross section is defined as
\begin{equation}
\tilde\sigma (N,M_H^2) = \int_0^1 d\tau_H \tau_H^{N-1} \sigma(s,M_H^2)
\end{equation}
so that the moments acquire the factorized form
\begin{equation}
\tilde \sigma(N-1,M_H^2) = \sigma_0 \sum_{ij} \tilde f_i(N,\mu_F^2)
\tilde f_j(N,\mu_F^2) \tilde G_{ij} \left(N;\alpha_s(\mu_R^2),
\frac{M_H^2}{\mu_R^2}; \frac{M_H^2}{\mu_F^2} \right)
\end{equation}
The Mellin transformation can be inverted by means of the contour
integral
\begin{equation}
\sigma(s,M_H^2) = \sigma_0 \sum_{ij} \int_{C-i\infty}^{C+i\infty}
\frac{dN}{2\pi i} \left(\frac{M_H^2}{s}\right)^{-N+1} \tilde f_i(N,\mu_F^2)
\tilde f_j(N,\mu_F^2) \tilde G_{ij} \left(N;\alpha_s(\mu_R^2),
\frac{M_H^2}{\mu_R^2}; \frac{M_H^2}{\mu_F^2} \right)
\end{equation}
where the value of the offset $C$ has to be chosen such that all
singularities of the $N$-moments are located on the left of the contour.
Soft and collinear gluon effects arise in the limit $z\to 1$ that
corresponds to the limit of large $N$ in Mellin space. The leading
contributions in Mellin space only appear in the $gg$ initial state,
while all other initial states are suppressed as ${\cal O}(1/N)$
\cite{gghnnll} as can be inferred explicitly from the NLO results
presented in Eq.~(\ref{eq:gghnlo})\footnote{Note that the soft and
collinear limits of the mass-dependent functions $d_{ij}^H(z,\tau_Q)$
coincide with the heavy top limits presented in Eq.~(\ref{eq:dijlim}).}.
Following Refs.~\cite{gghnnll,softresum} the leading contributions to
the coefficient function $\tilde G_{gg}$ in Mellin space are logarithmic
in $N$ and follow the perturbative expansion
\begin{eqnarray}
\tilde G_{gg} \left(N;\alpha_s(\mu_R^2),
\frac{M_H^2}{\mu_R^2}; \frac{M_H^2}{\mu_F^2} \right) \!\!\! & = &
\!\!\!\! \alpha_s^2(\mu_R^2) \left\{ 1 + \sum_{n=1}^\infty
\left(\frac{\alpha_s(\mu_R^2)}{\pi}\right)^n \sum_{m=0}^{2n} G_H^{(n,m)}
\log^m N \right\} + {\cal O}\left(\frac{1}{N} \right) \nonumber \\
& = & \tilde G_{gg}^{(res)} \left(N;\alpha_s(\mu_R^2),
\frac{M_H^2}{\mu_R^2}; \frac{M_H^2}{\mu_F^2} \right) + {\cal
O}\left(\frac{1}{N} \right)
\end{eqnarray}
where the second line indicates that the leading parts can be resummed
in closed form. The all-order resummed expression can be cast into the
form
\begin{eqnarray}
\tilde G_{gg}^{(res)} \left(N;\alpha_s(\mu_R^2),
\frac{M_H^2}{\mu_R^2}; \frac{M_H^2}{\mu_F^2}; \frac{M_H^2}{m_t^2} \right)
& = & \alpha_s^2(\mu_R^2) C_{gg} \left( \alpha_s(\mu_R^2),
\frac{M_H^2}{\mu_R^2}; \frac{M_H^2}{\mu_F^2}; \frac{M_H^2}{m_t^2} \right)
\nonumber \\
& & \times \exp \left\{ \tilde{\cal
G}_H \left( \alpha_s(\mu_R^2), \log N;
\frac{M_H^2}{\mu_R^2}, \frac{M_H^2}{\mu_F^2} \right) \right\}
\label{eq:gghres}
\end{eqnarray}
where we included top mass effects up to the NLL level explicitly as
will be discussed in following. The function $C_{gg}$ contains all
constant terms that originate from the $\delta(1-z)$ terms of the explicit
perturbative results and additional contributions emerging from the
Mellin transformation. It develops a perturbative expansion in the
strong coupling $\alpha_s$,
\begin{eqnarray}
C_{gg} \left( \alpha_s(\mu_R^2), \frac{M_H^2}{\mu_R^2};
\frac{M_H^2}{\mu_F^2}; \frac{M_H^2}{m_t^2} \right) & = & 1 + \sum_{n=1}^\infty
\left(\frac{\alpha_s(\mu_R^2)}{\pi}\right)^n C_{gg}^{(n)}
\left( \frac{M_H^2}{\mu_R^2}, \frac{M_H^2}{\mu_F^2}; \frac{M_H^2}{m_t^2} \right)
\end{eqnarray}
The leading logarithmic terms in $N$ are contained in the exponential
factor with the exponent
\begin{eqnarray}
\tilde{\cal G}_H \left( \alpha_s(\mu_R^2), \log N;
\frac{M_H^2}{\mu_R^2}, \frac{M_H^2}{\mu_F^2} \right) & = & \log
N~g_H^{(1)}(\lambda) \nonumber \\
& + & \left. \sum_{n=2}^\infty \alpha_s^{n-2}(\mu_R^2) g_H^{(n)}\left( \lambda,
\frac{M_H^2}{\mu_R^2}; \frac{M_H^2}{\mu_F^2} \right)
\right|_{\lambda=b_0 \alpha_s(\mu_R^2) \log N}
\label{eq:resum}
\end{eqnarray}
where $b_0$ denotes the leading order term of the QCD beta function,
\begin{equation}
b_0 = \frac{33-2N_F}{12\pi}
\end{equation}
where $N_F$ is the number of active flavours that we choose as
$N_F=5$ in the following, i.e.~the top quark has been decoupled from the
strong coupling $\alpha_s$ and the PDFs. The term containing the
function
\begin{equation}
g_H^{(1)}(\lambda) = \frac{3}{\pi b_0\lambda} \left[ 2\lambda +
(1-2\lambda)\log(1-2\lambda) \right]
\end{equation}
resums the leading logarithms in $N$, while the successive terms
with $g_H^{(i)}~(i\geq 2)$ cover the subleading,
subsubleading, etc. contributions. The explicit expressions for
$g_H^{(2)}$ and $g_H^{(3)}$ can be found in Refs.~\cite{gghnnll,g2g3}
and the one for $g_H^{(4)}$ in \cite{g4vogt} where the four-loop cusp
anomalous dimension has been approximated by employing Pad\'e
approximants. In the following we will use this approximate expression
for $g_H^{(4)}$. The NLO coefficient of the function $C_{gg}$ is given
by \cite{gghnnll}
\begin{eqnarray}
C_{gg}^{(1)} & = & \delta G_{gg}^{(1)} + 6(\gamma_E^2+\zeta_2) - 6\gamma_E
\log\frac{M_H^2}{\mu_F^2} \nonumber \\
\delta G_{gg}^{(1)} & = & c_H(\tau_t) + 6\zeta_2 + \frac{33-2N_F}{6}
\log\frac{\mu_R^2}{\mu_F^2}
\end{eqnarray}
where we included the top quark mass dependence explicitly by using the
function $c_H(\tau_t)$ \cite{gghnnllmt}. The NNLO coefficient
$C_{gg}^{(2)}$ in the limit of heavy top quarks can be found in
Ref.~\cite{gghnnll}. The N$^3$LO term can be extracted from the explicit
results given in \cite{gghn3loapprox} and can be found in Appendix E of
\cite{timo}.
In Ref.~\cite{limit} the inclusion of additional terms of purely
collinear origin in the resummation has been discussed and then
implemented in the results of \cite{gghnnlosv,gghnnll} by performing
the replacement
\begin{eqnarray}
C_{gg}^{(1)} & \to & C_{gg}^{(1)} + 6~\frac{\log N}{N}
\end{eqnarray}
This replacement reproduces all leading collinear logarithms of the form
$\alpha_s^n \log^{2n-1} N/N$ up to all orders. However, the subleading
logarithms of this type are not covered by this replacement as can be
inferred from the explicit NNLO expansion of the resummed expression in
Ref.~\cite{gghnnll}. We have observed that the subleading logarithms of
the type $\gamma_E \log^2 N/N$ and $\log M_H^2/\mu_F^2~\log^2 N/N$ can
be reproduced by the extended replacement
\begin{eqnarray}
C_{gg}^{(1)} & \to & C_{gg}^{(1)} + 6~\frac{\tilde L}{N}
\end{eqnarray}
with the modified logarithm
\begin{eqnarray}
\tilde L = \log \frac{N e^{\gamma_E} \mu_F}{M_H} = \log N + \gamma_E -
\frac{1}{2} \log \frac{M_H^2}{\mu_F^2}
\end{eqnarray}
This method can be extended to the next perturbative order. By performing
the NNLO replacement
\begin{eqnarray}
C_{gg}^{(2)} & \to & C_{gg}^{(2)} + (48-N_F)~\frac{\tilde L^2}{N}
\end{eqnarray}
we reproduce the correct collinear logarithms $\log^i N/N$ for $i=5,4$
at N$^3$LO,
\begin{eqnarray}
G_{gg}^{(3)} & = & 36 \log^6 N + 170.679 \log^5 N + 744.849 \log^4 N +
1405.185 \log^3 N \nonumber \\
& + & 2676.129 \log^2 N + 1897.141 \log N + 1783.692 \nonumber \\
& + & \frac{1}{N} \left\{ 108 \log^5 N + 615.696 \log^4 N \right\} + \ldots
\end{eqnarray}
where all numbers have been rounded at ${\cal O}(10^{-3})$ and the
additional contributions of the effective Lagrangian in the heavy top
limit have not been taken into account as in Ref.~\cite{gghn3loapprox}.
These terms agree with the threshold expansion of the N$^3$LO results of
\cite{gghn3loapprox} and the logarithmic scale dependence terms up to
the first two leading logarithms of ${\cal O}(1/N)$. A further inclusion
of subsubleading collinear logarithms requires a modified systematic
expansion of the Mellin transform of the resummed kernel in $\log \tilde
N = \log (N e^{\gamma_E})$ instead of $\log N$ and powers of $N^{-1}$
which is beyond the scope of this work. It should be noted that our
approach can be compared with the equivalent approach of
Ref.~\cite{mochvogt} that tries to construct a second exponential
resummation for the ${\cal O}(1/N)$ logarithms to be added to the
leading soft+virtual exponential. However, Ref.~\cite{mochvogt} does not
apply their conjecture to the Higgs boson case so that a direct
comparison of the methods is not possible presently.

The numerical impact of the subleading collinear logarithms and top mass
effects in the resummed expression ranges in the sub-per-mille range for
the SM Higgs mass. The inclusion of top mass effects in the resummation
can reach the per-cent level beyond NLO for large Higgs masses. An
alternative implementation of resummation effects includes the full
coefficient $C_{gg}$ of Eq.~(\ref{eq:gghres}) in the resummed
exponential with a careful expansion of the exponent up to all terms of
${\cal O}(1/N)$ included in the analysis along the lines of
Ref.~\cite{limit}. We have checked that this modification has an impact
in the per-mille range on the final results for the inclusive SM Higgs
cross sections.

\subsection{\label{sc:masseffects} Mass Effects and Matching}
Using the resummed expression of the gluon-fusion cross section in
Mellin space we subtract the corresponding fixed-order Mellin-space
result up to NNLO in order to obtain the net effect of resummation
beyond NNLO. This residual contribution has then been added to the NNLO
result that has been obtained by including the full NNLO result in the
limit of heavy top quarks for the pure top quark contributions and
adding top mass effects and bottom (charm) contributions at NLO
strictly. Furthermore we include only the top quark mass effects in the
resummed cross section at the NLL level (convolved with NLO $\alpha_s$
and PDFs) and treat the bottom- and charm-induced parts at fixed NLO. Since
the virtual coefficient of the bottom contributions
behaves in the limit $M_H^2\gg m_b^2$ as \cite{gghnlo} $(C_A = 3, C_F =
4/3)$
\begin{equation}
c_H (\tau_b) \to \frac{C_A-C_F}{12} \log^2 \frac{M_H^2}{m_b^2} - C_F \log
\frac{M_H^2}{m_b^2}
\end{equation}
if the bottom mass is renormalized on-shell, i.e.~it contains large
logarithms that are not resummed. The resummation of the Abelian part
proportional to $C_F$ has been performed in Ref.~\cite{hgagares} up to
the subleading logarithmic level. These logarithms are related to the
Sudakov form factor at the virtual $Hb\bar b$ vertex that generates
these large logarithmic contributions for far off-shell bottom quarks
inside the corresponding loop contributions in the Abelian case. The
resummation of the non-Abelian part proportional to the Casimir factor
$C_A$ has not been considered so far. This type of logarithmic
contributions emerges from a different origin than the soft and
collinear gluon effects discussed so far and is the main source of the
very different size of QCD corrections to the bottom-loop contributions
\cite{gghnlo,gghnlo0,gghnlosq0}. In order to obtain a reliable result
for the bottom contributions a resummation of these types of logarithms
is mandatory so that we do not include these contributions in our soft
and collinear gluon resummation but treat them at fixed NLO.

The complete cross section including soft/collinear gluon resummation
effects can then be cast into the generic form
\begin{eqnarray}
\!\!\!\!\!\!\! && \sigma(s,M_H^2) = \sigma_{tt}^0 \sum_{ij}
\int_{C-i\infty}^{C+i\infty} \frac{dN}{2\pi i}
\left(\frac{M_H^2}{s}\right)^{-N+1} \tilde f_g(N,\mu_F^2) \tilde
f_g(N,\mu_F^2) \nonumber \\
\!\!\!\!\!\!\! && \times \left\{ \tilde G^{(res)}_{gg}
\left(N;\alpha_s(\mu_R^2), \frac{M_H^2}{\mu_R^2};
\frac{M_H^2}{\mu_F^2};0 \right)
- \left[ \tilde G^{(res)}_{gg} \left(N;\alpha_s(\mu_R^2),
  \frac{M_H^2}{\mu_R^2}; \frac{M_H^2}{\mu_F^2};0 \right)
\right]_{(NNLO)} \right\} \nonumber \\
\!\!\!\!\!\!\! && + \sigma_{tt}^0 \sum_{ij} \int_{C-i\infty}^{C+i\infty}
\frac{dN}{2\pi i} \left(\frac{M_H^2}{s}\right)^{-N+1} \tilde
f_g(N,\mu_F^2) \tilde f_g(N,\mu_F^2) \nonumber \\
\!\!\!\!\!\!\! && \times \left\{ \tilde G^{(res)}_{gg,NLL}
\left(N;\alpha_s(\mu_R^2), \frac{M_H^2}{\mu_R^2}; \frac{M_H^2}{\mu_F^2};
\frac{M_H^2}{m_t^2} \right)
- \tilde G^{(res)}_{gg,NLL} \left(N;\alpha_s(\mu_R^2),
  \frac{M_H^2}{\mu_R^2}; \frac{M_H^2}{\mu_F^2};0 \right) \right.
\nonumber \\
\!\!\!\!\!\!\! && \left. - \left[\tilde G^{(res)}_{gg,NLL}
\left(N;\alpha_s(\mu_R^2), \frac{M_H^2}{\mu_R^2}; \frac{M_H^2}{\mu_F^2};
\frac{M_H^2}{m_t^2} \right)
- \tilde G^{(res)}_{gg,NLL} \left(N;\alpha_s(\mu_R^2),
  \frac{M_H^2}{\mu_R^2}; \frac{M_H^2}{\mu_F^2};0 \right) \right]_{(NLO)}
\right\} \nonumber \\
\!\!\!\!\!\!\! && + \sigma_{t+b+c}^{NNLO}(s,M_H^2)
\label{eq:resmatch}
\end{eqnarray}
with $\sigma_{tt}^0$ denoting the LO cross section factor $\sigma_0$ of
Eq.~(\ref{eq:ggh}) including only the top quark contribution. The index
'$(NNLO)$' in the second line indicates the fixed-order expansion of the
resummed coefficient function $\tilde G^{(res)}_{gg}$ in Mellin space up
to NNLO while the index '($NLO$)' denotes the perturbative expansion of
the NLL resummed coefficient function $\tilde G^{(res)}_{gg,NLL}$ in
Mellin space up to NLO. The first integral has been convolved with
N$^3$LO $\alpha_s$ and NNLO PDFs according to the discussion about the
non-necessity of N$^3$LO PDFs of Ref.~\cite{pdfn3lo} and
of resummed PDFs of Ref.~\cite{pdfresum} for the SM Higgs mass, while
the second integral has been evaluated with NLO $\alpha_s$ and PDFs
consistently. The fixed-order NNLO cross section of the last term has
been derived as
\begin{eqnarray}
\sigma_{t+b+c}^{NNLO}(s,M_H^2) = \sigma_{\infty}^{NNLO}(s,M_H^2) +
\sigma_{t+b+c}^{NLO}(s,M_H^2) - \sigma_{\infty}^{NLO}(s,M_H^2)
\end{eqnarray}
where the individual parts are defined as
\begin{eqnarray}
\sigma_{\infty}^{NNLO}(s,M_H^2) & = & \sigma_{tt}^{LO} K_\infty^{NNLO}
\nonumber \\
\sigma_{\infty}^{NLO}(s,M_H^2) & = & \sigma_{tt}^{LO} K_\infty^{NLO}
\nonumber \\
\sigma_{t+b+c}^{NLO}(s,M_H^2) & = & \sigma_{t+b+c}^{LO} K_{t+b+c}^{NLO}
\end{eqnarray}
where $\sigma_{tt}^{LO}$ denotes the full LO cross section including
only top loops, $\sigma_{t+b+c}^{LO}$ the LO cross section including top
and bottom/charm loops, $K_\infty^{(N)NLO}$ the (N)NLO K-factors
obtained in the limit of heavy top quarks and $K_{t+b+c}^{NLO}$ the full
NLO K-factor including top and bottom/charm loops.  The NNLO parts have
been derived with N$^3$LO $\alpha_s$ and NNLO PDFs and the NLO terms
with NLO $\alpha_s$ and PDFs consistently as implemented in the programs
HIGLU \cite{higlu} and SusHi \cite{sushi}. This implementation
guarantees that top mass effects are treated at NLL level and
bottom/charm contributions at fixed NLO respectively.

\subsection{Mellin Inversion}
As a final step we have to perform the Mellin inversion of the resummed
and properly matched result in Mellin space. According to
Eq.~(\ref{eq:resmatch}) this requires the inclusion of the Mellin
moments of the gluon densities $\tilde f_g(N,\mu_F^2)$ in the integrand
for the contour integration. This can be treated in different ways.
Refs.~\cite{gghnnll,gghnnllmt} have used a fit to the parton densities
in Bjorken-x space for a fixed factorization scale and used this fit to
determine the Mellin moments analytically in terms of a set of fitted
coefficients. We did not proceed along these lines. A second option will
be to use the evolution program {\tt PEGASUS} \cite{pegasus} in Mellin
space that just requires the implementation of the input densities in
terms of a predefined functional form. The latter does, however, not
coincide with the input densities of all available global PDF fits. We
have performed cross checks with {\tt PEGASUS} for the MSTW08 PDFs
\cite{mstw} at LO, NLO and NNLO\footnote{Differences between the
original MSTW08 evolution and the results of {\tt PEGASUS} at large
Bjorken-x have been clarified after including an updated set of MSTW08
PDFs \cite{timo}. The impact on the gluon-fusion cross section is small,
i.e.~at the per-mille level both for small and large Higgs masses, and
has been neglected in this analysis.}. The method that we adopted in our
numerical analysis is to implement the Mellin moments of the Bjorken-x
PDFs by a numerical integration. Due to the fact that the $z$-integral
of the fixed-order integral is extended to an infinite upper bound after
integration \cite{ttresum} due to the presence of the Landau singularity
at $N_L = \exp (1/[2 b_0 \alpha_s(\mu_R^2)])$ that implies that the
resummed kernel $\tilde G_{gg}^{(res)}$ does not vanish for $z>1$ but
drops down very fast for larger values of $z$. In order to increase the
numerical stability of the Mellin inversion we have included four
additional powers $1/(N-1)^4$ in the resummed kernels so that the large
$N$ contributions for $z>1$ are sufficiently suppressed. By means of
partial integrations this can be translated to the convolution over
derivatives of the gluon PDFs \cite{kulesza}\footnote{Note that the
translation to the second derivative appearing in the Mellin-integral
assumes that the gluon density and its first derivative vanish for
$x=1$.},
\begin{equation}
\tilde \sigma (N-1,M_H^2) = \sigma_0 \tilde{\cal F}_g(N,\mu_F^2)
\tilde{\cal F}_g(N,\mu_F^2) \tilde{\cal G}^{(res)}_{gg} \left(
N;\alpha_s(\mu_R^2), \frac{M_H^2}{\mu_R^2};
\frac{M_H^2}{\mu_F^2}\right)
\end{equation}
with
\begin{eqnarray}
\tilde{\cal G}^{(res)}_{gg} \left(
N;\alpha_s(\mu_R^2), \frac{M_H^2}{\mu_R^2};
\frac{M_H^2}{\mu_F^2}\right) & = & \tilde G^{(res)}_{gg} \left(
N;\alpha_s(\mu_R^2), \frac{M_H^2}{\mu_R^2};
\frac{M_H^2}{\mu_F^2}\right) / (N-1)^4 \nonumber \\
\tilde{\cal F}_g(N,\mu_F^2) & = & (N-1)^2 \tilde f_g(N,\mu_F^2)
\nonumber \\
& = & \int_0^1 dx x^{N-1} \frac{d}{dx}\left\{ x \frac{d}{dx} \left[ x
f_g(x,\mu_F^2) \right] \right\}
\end{eqnarray}
The second derivative has been implemented by a equidistant three-point
method
\begin{equation}
f''(x) = \frac{f(x+h)-2f(x)+f(x-h)}{h^2} + {\cal O}(h^2)
\end{equation}
Using this method we can implement the original PDF fits in Bjorken-$x$
space and obtain sufficient numerical stability if the contour is chosen
according to the parametrization \cite{kulesza}
\begin{equation}
N = C + x e^{\pm i\phi}
\end{equation}
with the $+(-)$ sign applied to the upper (lower) contour with respect
to the real axis and the integration proceeds over the variable $x$. The
off-set parameter has been chosen as $C=2.5$ and the angle $\phi$ as
$3\pi/4$ for $z<1$ and $\pi/4$ for $z>1$. This choice ensures that all
relevant singularities in Mellin space are located to the left of the
integration contour, but the Landau singularity to the right, i.e.~the
latter is excluded from the contour integral according to the minimal
prescrition method of Ref.~\cite{ttresum}. We have checked the
independence of our results of variations of the two parameters $C,\phi$
around the chosen values within the valid ranges, i.e.~keeping $C$ in
the range $2<C<N_L$ and the signs of the angle $\phi$ and its
hemispheres with respect to $\pi/2$.

\section{MSSM Higgs Boson Production}
In the MSSM all three neutral Higgs bosons are produced via gluon fusion
$gg\to h,H,A$. In the following we will neglect the stop and sbottom
loops in MSSM scalar Higgs boson production and will focus on the top
and bottom-induced contributions that are modified in the scalar case
due to the additional mixing factors of the top and bottom Yukawa
couplings compared to the SM Higgs case as given in Table \ref{tb:hcoup}.
This approximation works for heavy stops and sbottoms. Pseudoscalar
Higgs boson production, however, requires a transformation of the SM
Higgs results to the production of a ${\cal CP}$-odd Higgs particle. The
resummed exponential of Eq.~(\ref{eq:gghres}) is universal and thus
unchanged in the pseudoscalar case (if $M_H$ is replaced by the
pseudoscalar Higgs mass $M_A$) so that only the coefficient $C_{gg}$ is
different from scalar Higgs boson production. The difference of this
coefficient between the pseudoscalar and scalar Higgs cases is given
by\footnote{Note that the NNLO expression differs from
Ref.~\cite{ggannll}. The (numerically minor) discrepancies have been
clarified with the authors.}
\begin{eqnarray}
\Delta C_{gg}^{(1)} & = & C_{gg,A}^{(1)} - C_{gg,H}^{(1)} = c_A(\tau_t)
- c_H(\tau_t) \nonumber \\
\Delta C_{gg}^{(2)} & = & C_{gg,A}^{(2)} - C_{gg,H}^{(2)} \nonumber \\
& = & \frac{1939}{144} + 3\gamma_E^2 + 6\zeta_2 -\frac{21}{16} N_F -
\left(\frac{19}{8}-\frac{N_F}{3} \right) \log\frac{M_A^2}{m_t^2} \nonumber \\
& + & \left( \frac{33-2N_F}{12} - 3\gamma_E \right)
\log\frac{M_A^2}{\mu_F^2} - \frac{33-2N_F}{8} \log\frac{M_A^2}{\mu_R^2}
\end{eqnarray}
where at NLO we again include the full top mass dependence in this
coefficient along the lines of Ref.~\cite{gghnnllmt}. In the limit of
heavy top quarks the difference of the virtual corrections at NLO
approaches
\begin{equation}
\Delta C_{gg}^{(1)} = C_{gg,A}^{(1)} - C_{gg,H}^{(1)} \to \frac{1}{2}
\end{equation}
Since the gluon-fusion production cross section for pseudoscalar Higgs
bosons is only known up to NNLO we reduce our resummation to the NNLL
level, i.e.~we do not include the function $g_H^{(4)}$ in the resummed
exponent of Eq.~(\ref{eq:resum}). The inclusion of purely collinear
logarithms proceeds along the same lines as for the SM Higgs case by
implementing the replacements
\begin{eqnarray}
C_{gg,A}^{(1)} & \to & C_{gg,A}^{(1)} + 6 \frac{\tilde L}{N} \nonumber \\
C_{gg,A}^{(2)} & \to & C_{gg,A}^{(2)} + (48-N_F) \frac{\tilde L^2}{N}
\end{eqnarray}
with the extended logarithm
\begin{eqnarray}
\tilde L = \log \frac{N e^{\gamma_E} \mu_F}{M_H} = \log N + \gamma_E -
\frac{1}{2} \log \frac{M_A^2}{\mu_F^2}
\end{eqnarray}
These replacements resum the corresponding collinear logarithms at
leading and subleading level accordingly.

\section{Results}
We analyze Higgs boson production via gluon fusion at the LHC for a
c.m.~energy of $\sqrt{s}=13$ TeV for the SM Higgs boson and the neutral
Higgs bosons of the MSSM at the N$^3$LL level for the top-induced
contributions of the scalar Higgs cases and at the NNLL level for the
pseudoscalar case. As parton density functions (PDFs) we will primarily
use the MSTW08 sets \cite{mstw} with the strong coupling normalized to
$\alpha_s (M_Z^2) = 0.12018$ at NLO and to $\alpha_s (M_Z^2) = 0.11707$
(N)NNLO. The quark pole masses have been chosen as $m_t = 172.5$ GeV,
$m_b = 4.75$ GeV and $m_c = 1.40$ GeV accordingly. For the determination
of the PDF+$\alpha_s$ uncertainties we have used the envelope method
\cite{hxswg} with CT10 \cite{ct10} and NNPDF2.3
\cite{nnpdf} PDF sets. For CT10 the strong coupling has been normalized
consistently to $\alpha_s=0.118$ and for NNPDF2.3 to $\alpha_s=0.119$,
respectively. The uncertainty in the strong coupling
constant has been adopted as $\pm 0.002$ around the corresponding
central values of the used PDFs\footnote{If other PDF sets as ABM12
\cite{abm} or HERAPDF1.5 \cite{herapdf} are included in this envelope
the PDF+$\alpha_s$ uncertainties will increase considerably with a major
part originating from sizeable differences in the $\alpha_s$ fits at
NNLO and different data sets included in the global fits. Moreover, the
proper treatment of higher-twist effects in the global fits is an open
aspect in this context. However, we have adopted the scheme used within
the HXSWG, since an extended study of this particular issue is beyond
the scope of our paper.}.

\subsection{Standard Model}
\begin{figure}[hbt]
\begin{picture}(200,270)(0,0)
\put(0,-170){\includegraphics{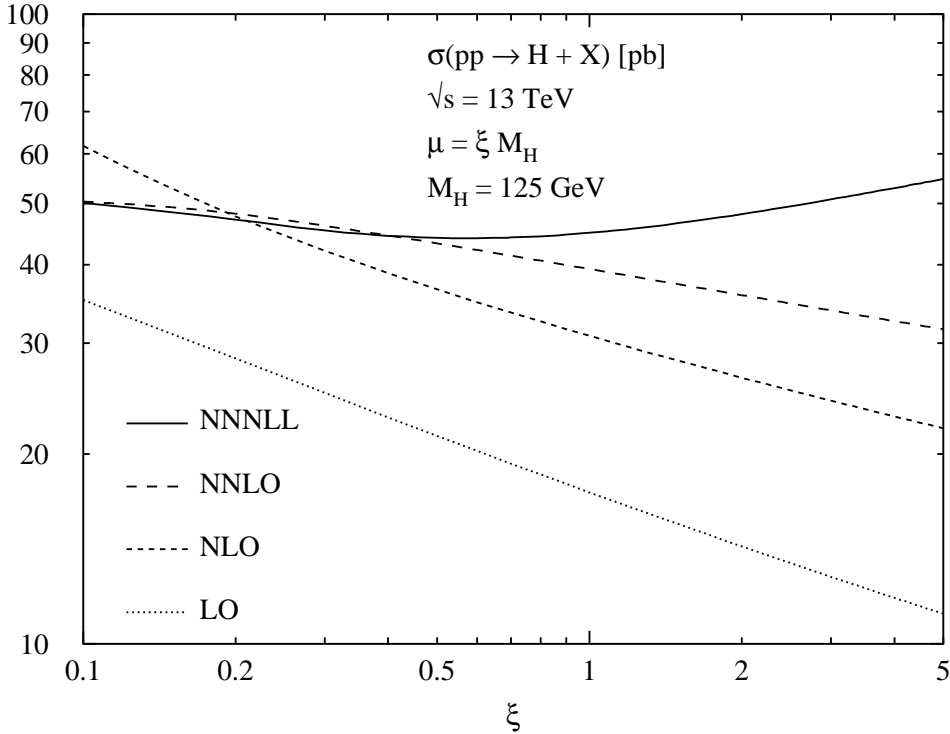}}
\end{picture}
\caption{\it \label{fg:scale} Scale dependence of the SM Higgs
production cross sections via gluon fusion for a Higgs mass $M_H=125$
GeV and c.m.~energy $\sqrt{s}=13$ TeV using MSTW08 PDFs. The
renormalization and factorization scales are identified,
i.e.~$\mu=\mu_R=\mu_F$.}
\end{figure}
For the SM Higgs boson we perform the numerical analysis for Higgs mass
values around the measured Higgs mass of about 125 GeV. For the
uncertainties we consider the scale dependence and the PDF+$\alpha_s$
uncertainties. The scale dependence at LO, NLO, NNLO and N$^3$LL is
shown in Fig.~\ref{fg:scale} as a function of the identified
renormalization and factorization scales in units of the Higgs mass
$M_H$. We observe a significant decrease of the scale dependence with
increasing perturbative order signalizing a rather mild scale dependence
at N$^3$LL. However, this is caused by a significant cancellation
between the renormalization and factorization scale dependences that
develop opposite slopes \cite{gghnnll}.  For a consistent estimate of
the theoretical uncertainties both scales have to be varied
independently. The results of varying both scales by a factor of 2 up
and down around our central scale choice $\mu_0 = M_H/2$ are shown in
Table \ref{tb:scale} without electroweak corrections. The numbers
develop a much larger variation within the complete interval of
independent variations so that identifying the renormalization and
factorization scales leads to a significant underestimate of the related
uncertainty. Taking the minimal and maximal values of Table
\ref{tb:scale} excluding the values for $\mu_F/\mu_R > 2$ and
$\mu_F/\mu_R < \frac{1}{2}$ we derive a total scale dependence of
$[+4.4\%, -5.3\%]$, while for identified scales we
obtain the optimistic estimate of $[+4.4\%, -0.1\%]$. Thus for a
sophisticated determination of the theoretical uncertainties due to the
scale dependence both scales have to be varied independently contrary to
what has been done in Ref.~\cite{gghn3lo}.
\begin{table}[hbt]
\renewcommand{\arraystretch}{1.5}
\begin{center}
\begin{tabular}{|c||c|c|c|} \hline
\backslashbox[2cm]{$\mu_F/M_H$}{$\mu_R/M_H$} & 1/4 & 1/2 &  1 \\
\hline \hline
1/4 & $46.08~pb$ & $42.92~pb$ & $39.35~pb$ \\ \hline
1/2 & $45.42~pb$ & $44.13~pb$ & $41.81~pb$ \\ \hline
1   & $44.98~pb$ & $45.81~pb$ & $44.94~pb$ \\ \hline
\end{tabular}
\renewcommand{\arraystretch}{1.2}
\caption{\label{tb:scale} \it N$^{\,3\!}$LL Higgs boson production cross
sections via gluon fusion for different values of the renormalization
and factorization scales $\mu_R, \mu_F$ without electroweak corrections
for a Higgs mass $M_H=125$ GeV and c.m.~energy $\sqrt{s}=13$ TeV.}
\end{center}
\end{table}

Including electroweak corrections \cite{gghelw} we present our
predictions of the central cross sections for five different Higgs mass
values in Table \ref{tb:mstw} where we choose the MSTW08 PDFs
\cite{mstw} and add the corresponding PDF+$\alpha_s$ uncertainties. The
total uncertainty is obtained by adding the scale and PDF+$\alpha_s$
uncertainties linearly. In this way the total uncertainties amount to
${\cal O}(10\%)$. Additional uncertainties due to parametric
uncertainties in the top and bottom/charm quark masses and missing quark
mass effects beyond NLO are small, i.e.~in the per-cent range and thus
negligible.
\begin{table}[hbt]
\renewcommand{\arraystretch}{1.5}
\begin{center}
\begin{tabular}{|c||c|c|c|c|} \hline
$M_H$ [GeV] & $\sigma(pp \to H + X)$ [pb] & scale & PDF+$\alpha_s$ &
total \\ \hline \hline
124   & $47.05~pb$ & $^{+4.5\%}_{-5.2\%}$ & $^{+3.7\%}_{-4.0\%}$
& $^{+8.2\%}_{-9.2\%}$ \\ \hline
124.5 & $46.72~pb$ & $^{+4.4\%}_{-5.3\%}$ & $^{+3.7\%}_{-4.0\%}$ &
$^{+8.1\%}_{-9.3\%}$ \\ \hline
125   & $46.40~pb$ & $^{+4.4\%}_{-5.3\%}$ & $^{+3.7\%}_{-4.0\%}$
& $^{+8.1\%}_{-9.3\%}$ \\ \hline
125.5 & $46.06~pb$ & $^{+4.4\%}_{-5.2\%}$ & $^{+3.7\%}_{-4.0\%}$ &
$^{+8.1\%}_{-9.2\%}$ \\ \hline
126   & $45.74~pb$ & $^{+4.4\%}_{-5.2\%}$ & $^{+3.7\%}_{-4.0\%}$
& $^{+8.1\%}_{-9.2\%}$ \\ \hline
\end{tabular}
\renewcommand{\arraystretch}{1.2}
\caption{\label{tb:mstw} \it N$^{\,3\!}$LL Higgs boson production cross
sections via gluon fusion for different values of the Higgs mass
including the individual uncertainties due to the renormalization and
factorization scale dependence and PDF+$\alpha_s$ uncertainties
including electroweak corrections using MSTW08 PDFs for a c.m.~energy
$\sqrt{s}=13$ TeV.}
\end{center}
\end{table}
In Table \ref{tb:cxn} we present our final predictions of the gluon
fusion cross section for five different Higgs masses with the central
values and PDF+$\alpha_s$ obtained by the envelope method
\cite{hxswg} involving MSTW08 \cite{mstw}, CT10
\cite{ct10} and NNPDF2.3 \cite{nnpdf} PDFs.  This results in a slight
increase of the total uncertainties to the level of ${\cal O}(15\%)$.
This situation will improve with the inclusion of more recent PDFs
including LHC data in their global fits.
\begin{table}[hbt]
\renewcommand{\arraystretch}{1.5}
\begin{center}
\begin{tabular}{|c||c|c|c|c|}
\hline $M_H$ [GeV] & $\sigma(pp \to H + X)$ [pb] & scale & PDF+$\alpha_s$
& total \\ \hline \hline
124   & $48.37~pb$ & $^{+4.5\%}_{-5.2\%}$ & $\pm 8.1\%$ &
$^{+12.6\%}_{-13.3\%}$ \\ \hline
124.5 & $48.00~pb$ & $^{+4.4\%}_{-5.3\%}$ & $\pm 8.1\%$ &
$^{+12.5\%}_{-13.4\%}$ \\ \hline
125   & $47.63~pb$ & $^{+4.4\%}_{-5.3\%}$ & $\pm 8.1\%$ &
$^{+12.5\%}_{-13.4\%}$ \\ \hline
125.5 & $47.28~pb$ & $^{+4.4\%}_{-5.2\%}$ & $\pm 8.2\%$ &
$^{+12.6\%}_{-13.4\%}$ \\ \hline
126   & $46.94~pb$ & $^{+4.4\%}_{-5.2\%}$ & $\pm 8.3\%$ &
$^{+12.7\%}_{-13.5\%}$ \\ \hline
\end{tabular}
\renewcommand{\arraystretch}{1.2}
\caption{\label{tb:cxn} \it N$^{\,3\!}$LL Higgs boson production cross
sections via gluon fusion for different values of the Higgs mass
including the individual and total uncertainties due to the
renormalization and factorization scale dependence and PDF+$\alpha_s$
uncertainties including electroweak corrections using the envelope of
MSTW08 \cite{mstw}, CT10 \cite{ct10} and NNPDF2.3 \cite{nnpdf} PDFs for
a c.m.~energy $\sqrt{s}=13$ TeV.}
\end{center}
\end{table}
\begin{table}[hbt]
\renewcommand{\arraystretch}{1.5} \begin{center}
\begin{tabular}{|c||c|c|c|c|c|} \hline $\mu/M_H$ & Ref.~\cite{gghn3lo} &
SVC$_\infty$ & N$^3$LL$_\infty$ & massive N$^3$LL & with elw.  corr. \\
\hline \hline
1/2 & $44.31^{+0.3\%}_{-2.6\%}~pb$ & $44.22^{+4.2\%}_{-1.0\%}~pb$ &
$44.15^{+4.6\%}_{-0.1\%}~pb$ & $44.13^{+4.4\%}_{-0.1\%}~pb$ &
$46.40^{+4.4\%}_{-0.1\%}~pb$ \\ \hline
1   & $43.14^{+2.7\%}_{-4.5\%}~pb$ & $43.77^{+1.0\%}_{-0.0\%}~pb$ &
$44.72^{+6.3\%}_{-1.3\%}~pb$ & $44.94^{+6.6\%}_{-1.8\%}~pb$ &
$47.25^{+6.6\%}_{-1.8\%}~pb$ \\ \hline
\end{tabular}
\renewcommand{\arraystretch}{1.2}
\caption{\label{tb:comparison} \it Higgs boson production cross sections
via gluon fusion for different values of the identified renormalization
and factorization scales $\mu = \mu_R =\mu_F$ with and without quark
mass effects and electroweak corrections in comparison to the
N$^{\,3\!}$LO results of Ref.~\cite{gghn3lo} using MSTW08 PDFs for a
c.m.~energy $\sqrt{s}=13$ TeV. The percentage errors provide the
relative scale dependence for identified renormalization and
factorization scales $\mu = \mu_R =\mu_F$ varied by a factor of 2 up and
down around the central scale choice. The second and third columns of
cross sections in the heavy top limit at all perturbative orders,
labelled 'SVC$_\infty$' and 'N$^{\,3\!}$LL$_\infty$', exclude
bottom/charm contributions.}
\end{center}
\end{table}

Finally, we compare our results with the N$^3$LO predictions of
Ref.~\cite{gghn3lo} in Table \ref{tb:comparison}.  This table shows the
numbers of Ref.~\cite{gghn3lo} in the first column which have been
obtained in the limit of heavy top quarks for the QCD corrected cross
sections at all perturbative orders and neglecting the bottom and charm
loops. The renormalization and factorization scales have been identified
for the derivation of the scale dependence given in per-cent attached to
each number. The second column labelled 'SVC$_\infty$' depicts our
prediction in the heavy top limit by using the approximate
soft+virtual+collinear expansion of our resummed kernel for the N$^3$LO
piece added to the full NNLO result in the same limit. These approximate
fixed-order results agree with the explicit numbers of
Ref.~\cite{gghn3lo} within 0.2\% for $\mu_R=\mu_F=M_H/2$ and
within 1.5\% for $\mu_R=\mu_F=M_H$ and develop a similar scale
dependence. Our corresponding result at N$^3$LL is shown in the third
column labelled 'N$^3$LL$_\infty$' in the limit of heavy top quarks and
omitting the bottom and charm contributions. Our scale dependence for
identified scales is of similar size as the one of the N$^3$LO results.
However, this scale dependence does not constitute a reliable estimate
of the theoretical uncertainties as discussed before.  The comparison of
the second and third column shows the effect of resummation beyond
N$^3$LO.  It is clearly visible that the resummation effects range at
the per-mille level for the scale choice $\mu_R=\mu_F=M_H/2$, while
resummation provides a 2\% contribution beyond N$^3$LO for the scale
choice $\mu_R=\mu_F=M_H$. The latter effect is of the order of the
uncertainties related to the scale dependence.  The fourth column
presents our full N$^3$LL results including top mass effects up to NLL
and bottom/charm loops up to NLO.  The value for our central scale
choice $\mu_R=\mu_F=M_H/2$ agrees with the number of Ref.~\cite{gghn3lo}
within 0.3\% by accident. The last column shows our numbers including
the electroweak corrections of Ref.~\cite{gghelw}. Our final results
deviate by a few per cent from the numbers presented in
\cite{gghn3ll1,gghn3ll2}, since the first analysis works at approximate
fixed N$^3$LO and does neither include top mass effects beyond LO nor
bottom/charm loops and in both analyses electroweak corrections have not
been taken into account. Moreover, small differences can arise due to
different implementations of the PDFs in the resummation framework in
comparison with the second analysis \cite{gghn3ll2}.

\subsection{Minimal supersymmetric extension}
For neutral MSSM Higgs boson production via gluon fusion we have
adopted the $m_h^{mod+}$ scenario of Ref.~\cite{benchmark} which is
defined as
\begin{eqnarray}
M_{SUSY} = 1~{\rm TeV}, \mu = M_2 = 200~{\rm GeV}, X_t = 1.6 M_{SUSY},
m_{\tilde g} = 1.5~{\rm TeV}, A_b = A_\tau = A_t
\end{eqnarray}
where $M_{SUSY}$ denotes the SUSY-breaking sfermion mass scale of the
third generation, $\mu$ the higgsino mass parameter, $M_2$ the wino mass
parameter, $X_t = A_t - \mu \tgb$ the stop mixing parameter, $m_{\tilde
g}$ the gluino mass and $A_{t,b,\tau}$ the trilinear SUSY-breaking
couplings of the sfermions and Higgs fields.  We use the RG-improved
two-loop expressions for the Higgs masses and couplings of
Ref.~\cite{rgi} which yield predictions for the Higgs boson masses that
agree with the diagrammatic calculations of Ref.~\cite{mssmrad} within
3--4\% in general. Thus the leading one- and two-loop corrections have
been included in the Higgs masses and the effective mixing angle
$\alpha$.

\begin{figure}[hbtp]
\begin{picture}(200,250)(0,0)
\put(0,-170){\includegraphics{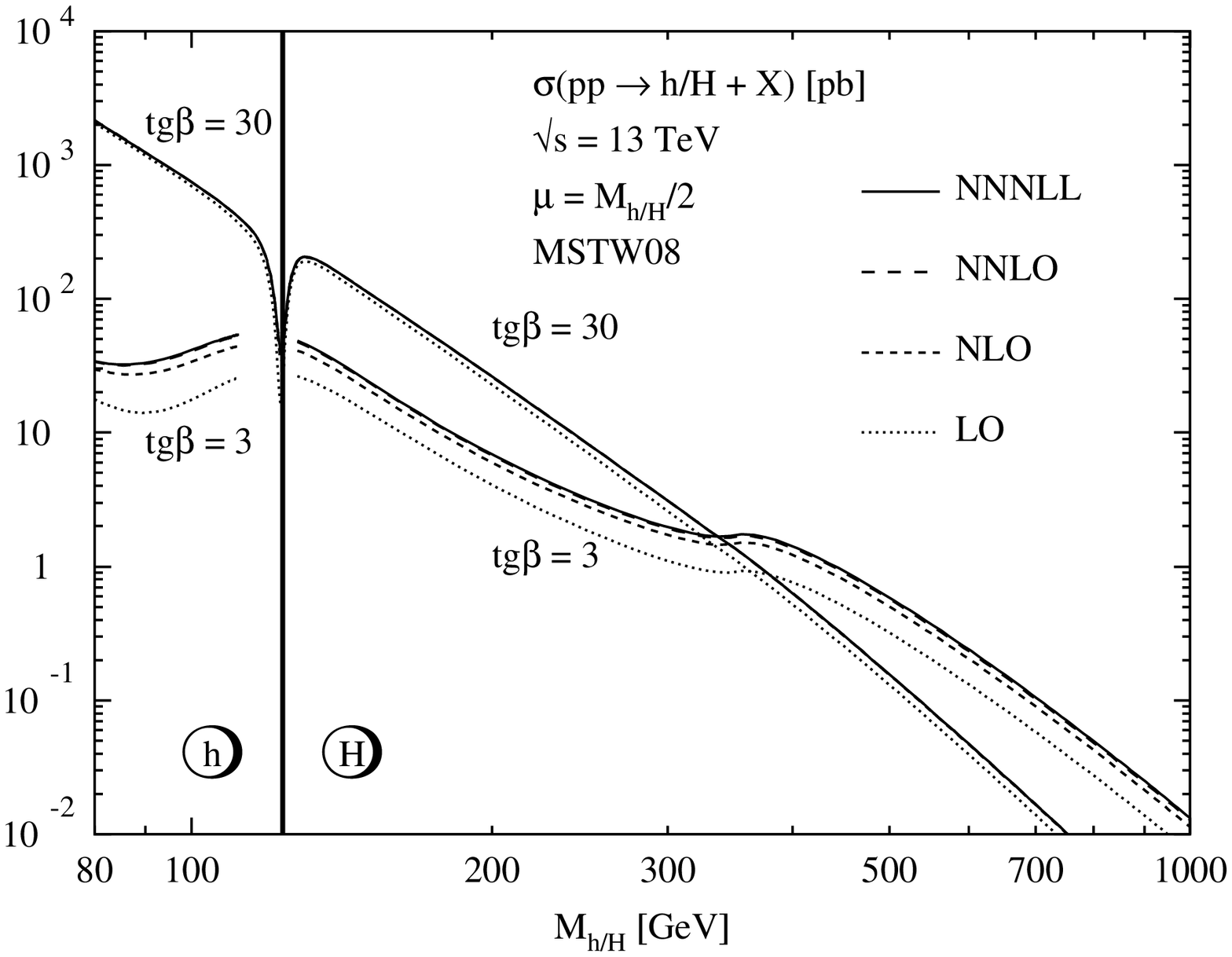}}
\put(0,-450){\includegraphics{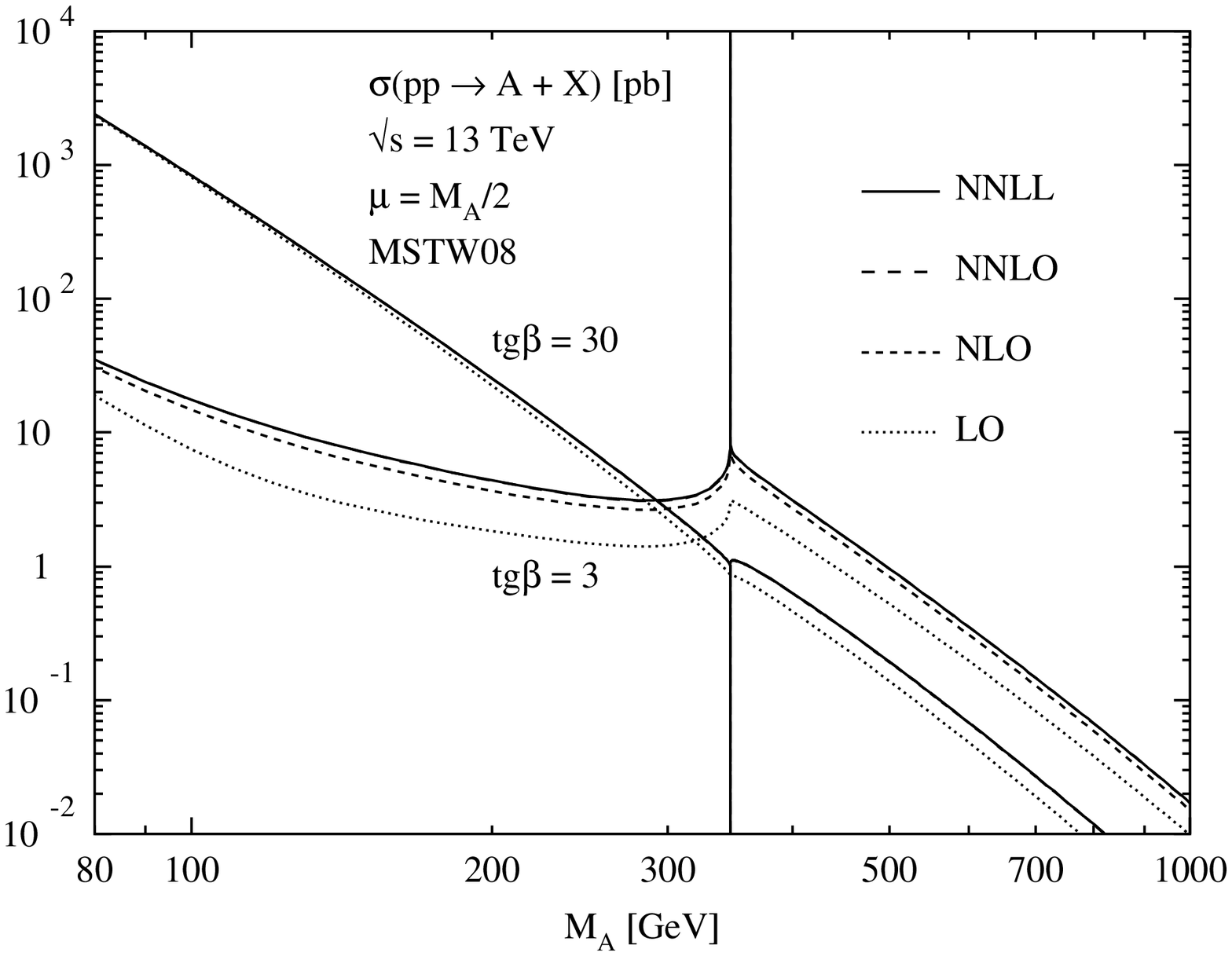}}
\end{picture}
\vspace*{10cm}

\caption{\it \label{fg:cxnmssm} MSSM Higgs production cross sections via
gluon fusion at the LHC for a c.m.~energy $\sqrt{s}=13$ TeV and two
values of $\tgb=3,30$ using MSTW08 PDFs. For the MSSM the $m_h^{mod+}$
scanario \cite{benchmark} has been adopted and squark loops as well as
genuine SUSY--QCD corrections have been neglected.}
\end{figure}

\begin{figure}[hbtp]
\begin{picture}(200,250)(0,0)
\put(0,-170){\includegraphics{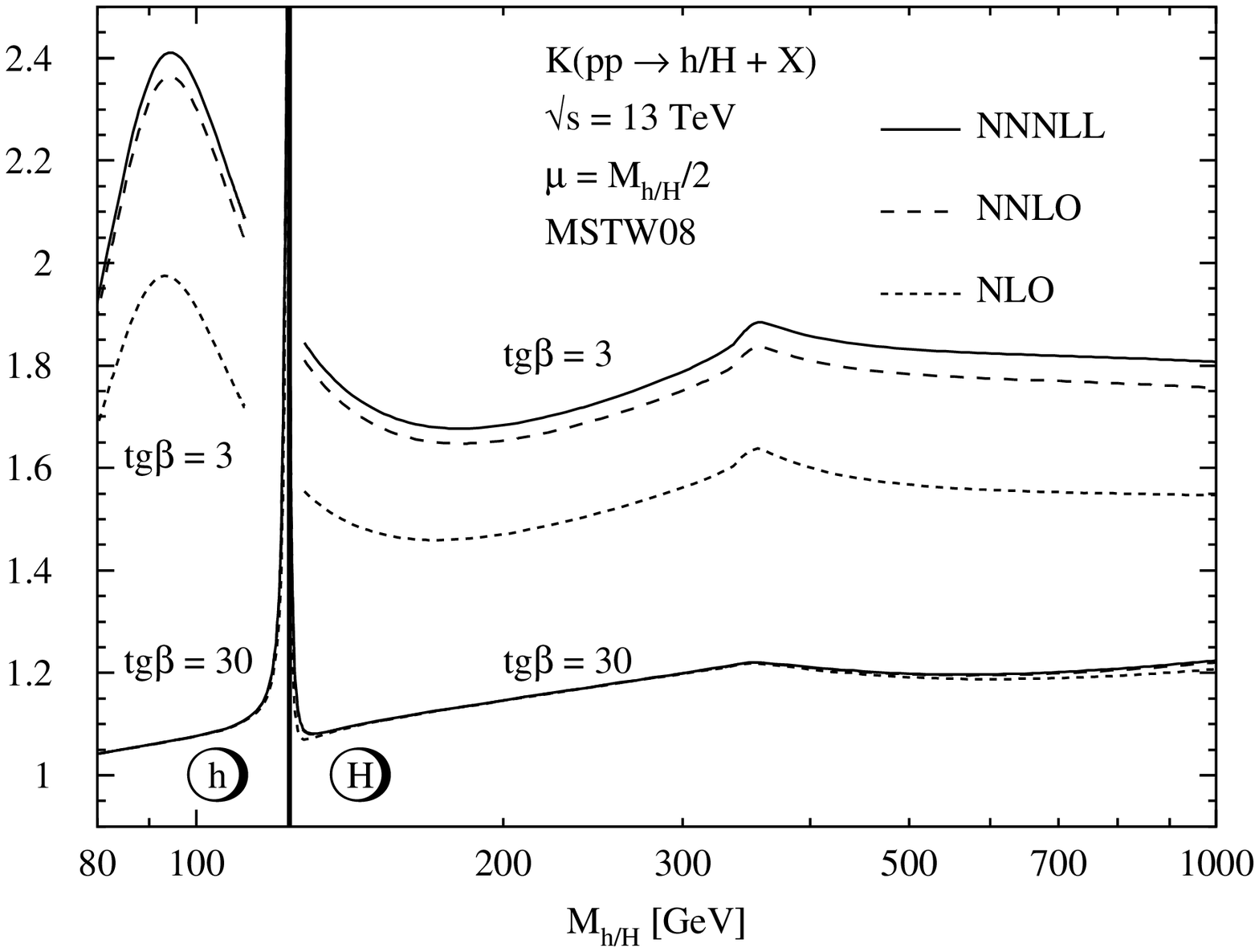}}
\put(0,-450){\includegraphics{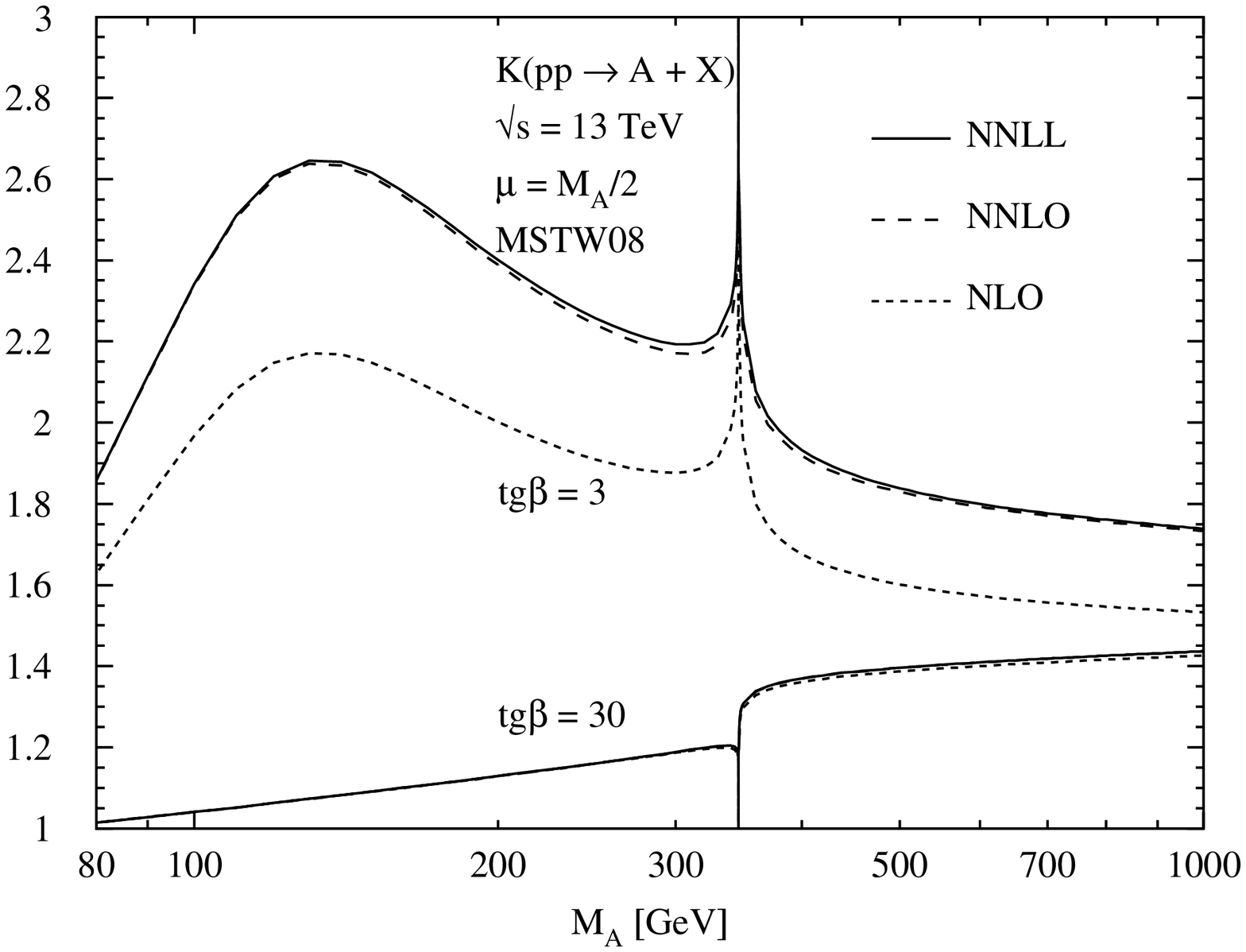}}
\end{picture}
\vspace*{10cm}

\caption{\it \label{fg:kmssm} K-factors of the MSSM Higgs production
cross sections via gluon fusion at the LHC for a c.m.~energy
$\sqrt{s}=13$ TeV and two values of $\tgb=3,30$ using MSTW08 PDFs. For
the MSSM the $m_h^{mod+}$ scenario \cite{benchmark} has been adopted and
squark loops as well as genuine SUSY--QCD corrections have been
neglected.}
\end{figure}

In Fig.~\ref{fg:cxnmssm} we show the scalar and pseudoscalar production
cross sections as functions of the corresponding Higgs masses at LO,
NLO, NNLO and (N)NNLL for the pseudoscalar (scalar) MSSM Higgs bosons.
Squark loops and genuine SUSY--QCD corrections have been neglected in
this work. They can be added in a form factorized from the resummation
effects which is left for future work. The resummation effects beyond
NNLO amount to about 5\% for $\tgb=3$, where the top loop contributions
are dominant, while for large values of $\tgb=30$ the effects are small,
since the bottom loops dominate so that the accuracy of the cross
section is of fixed NLO according to our setup as discussed in Section
\ref{sc:masseffects}. This can be inferred more clearly from
Fig.~\ref{fg:kmssm} which shows the corresponding K-factors for scalar
and pseudoscalar Higgs boson production via gluon fusion. The K-factors
are defined as the ratios of the NLO, NNLO and (N)NNLL cross
sections to the LO prediction where each order is conistently evaluated
with the corresponding PDF and $\alpha_s$ choices. For large values of
$\tgb=30$ it is obvious from this figure that there is only a tiny
effect beyond NLO as expected due to the dominance of the bottom loops
that are purely treated at fixed NLO. The bumps and spikes for $M_{H/A}\sim
2m_t$ are related to the $t\bar t$ threshold that generates a Coulomb
singularity for the pseudoscalar case already at NLO \cite{gghnlo,gganlo}.
The latter is regularized by taking into account the finite width of the
virtual top quarks which, however, is beyond the scope of our work. The
results indicate that the QCD uncertainties reduce to the level of $\sim 10\%$
after including the resummation effects in regions of top-loop dominance.
The dominant uncertainties in these regions within the MSSM will arise
from genuine SUSY effects.

\section{Conclusions}
In this work we have performed an analysis of soft and collinear gluon
resummation effects in SM and MSSM Higgs boson production via gluon
fusion at the LHC. For scalar Higgs boson production the results have
been analyzed at the N$^3$LL level, while pseudoscalar Higgs boson
production can only be extended to the NNLL level with the present
state-of-the-art calculations. We have implemented top mass effects up
to the NLL level by using the full mass dependence of the finite part of
the virtual corrections according to \cite{gghnnllmt}. The bottom and
charm loops, however, have been treated at fixed NLO rigorously since
there are additional non-resummed double logarithmic contributions that
have a large impact on the size of their contribution. We have discussed
an alternative extension of the formerly used approach for the inclusion
of collinear gluon effects at ${\cal O}(1/N)$ in Mellin space to the
subleading logarithmic level as a conjecture\footnote{The recently
developed next-to-eikonal approach \cite{nte} may provide a basis for a
factorization proof of logarithmic ${\cal O}(1/N)$ terms.}. We have been
able to reproduce the leading and subleading N$^3$LO terms at ${\cal
O}(1/N)$ which is a non-trivial cross check of our method. The impact of
resummation effects on the total cross sections reaches a size of a few
per cent beyond the fixed-order calculations, while the effect of
including top mass effects and subleading collinear logarithms in the
resummation ranges at the per-mille level or below. \\

\noindent
{\bf Acknowledgements.} We are indebted to C.~Anastasiou, J.~Bl\"umlein,
D.~de Florian, M.~Grazzini, S.~Moch, J.~Rojo, R.~Thorne, A.~Vogt and
J.~Zurita for useful discussions and helpful clarifications.  We are
grateful to R.~Thorne for providing us with an updated version of the
MSTW08 PDFs.  Ths work is supported in part by the Research Executive
Agency (REA) of the European Union under the Grant Agreement
PITN-GA-2012-316704 (Higgstools).

\end{document}